%% file: 0_main.tex
\documentclass[conference]{IEEEtran}
\IEEEoverridecommandlockouts
\usepackage{cite}
\usepackage{amsmath,amssymb,amsfonts}
\usepackage{algorithmic}
\usepackage{graphicx}
\usepackage{textcomp}
\usepackage{siunitx}
\usepackage{xcolor}
\usepackage{float}
\usepackage{dblfloatfix}
\usepackage{soul}
\usepackage{cancel} % for \sout usage (strikethrough)
\usepackage{tabularx}
\usepackage{array}
\usepackage{bm}
\usepackage[section]{placeins}

\usepackage{tikz}
\usepackage{textcomp}
\usepackage{lipsum}

\def\BibTeX{{\rm B\kern-.05em{\sc i\kern-.025em b}\kern-.08em
    T\kern-.1667em\lower.7ex\hbox{E}\kern-.125emX}}

\newcommand\copyrighttext{%
  \footnotesize \textcopyright 2023 IEEE. Personal use of this material is permitted.
  Permission from IEEE must be obtained for all other uses, in any current or future
  media, including reprinting/republishing this material for advertising or promotional
  purposes, creating new collective works, for resale or redistribution to servers or
  lists, or reuse of any copyrighted component of this work in other works.
  DOI: 10.1109/ICCAD57390.2023.10323763.}
\newcommand\copyrightnotice{%
\begin{tikzpicture}[remember picture,overlay]
\node[anchor=south,yshift=10pt] at (current page.south) {\fbox{\parbox{\dimexpr\textwidth-\fboxsep-\fboxrule\relax}{\copyrighttext}}};
\end{tikzpicture}%
}
\begin{document}

\title{Analog or Digital In-memory Computing? Benchmarking through Quantitative Modeling}
  
\author{
\IEEEauthorblockN{Jiacong Sun}
\IEEEauthorblockA{\textit{MICAS, KU Leuven}\\
Leuven, Belgium \\
jiacong.sun@kuleuven.be}
\and 
\IEEEauthorblockN{Pouya Houshmand}
\IEEEauthorblockA{\textit{MICAS, KU Leuven} \\
Leuven, Belgium \\
pouya.houshmand@kuleuven.be}
\and
\IEEEauthorblockN{Marian Verhelst}
\IEEEauthorblockA{\textit{MICAS, KU Leuven} \\
Leuven, Belgium\\
marian.verhelst@kuleuven.be}
}

\maketitle
\copyrightnotice

\begin{abstract}
In-Memory Computing (IMC) has emerged as a promising paradigm for energy-efficient, throughput-efficient and area-efficient machine learning at the edge. However, the differences in hardware architectures, array dimensions, and fabrication technologies among published IMC realizations have made it difficult to grasp their relative strengths. Moreover, previous studies have primarily focused on exploring and benchmarking the peak performance of a single IMC macro rather than full system performance on real workloads. This paper aims to address the lack of a quantitative comparison of Analog In-Memory Computing (AIMC) and Digital In-Memory Computing (DIMC) processor architectures. We propose an analytical IMC performance model that is validated against published implementations and integrated into a system-level exploration framework for comprehensive performance assessments on different workloads with varying IMC configurations. 
Our experiments show that while DIMC generally has higher computational density than AIMC, AIMC with large macro sizes may have better energy efficiency than DIMC on convolutional-layers and pointwise-layers, which can exploit high spatial unrolling. On the other hand, DIMC with small macro size outperforms AIMC on depthwise-layers, which feature limited spatial unrolling opportunities inside a macro. 

\end{abstract}


\input{1_introduction}

\input{2_background}

\input{4_imc_model}
\input{5_model_validation}

\input{6_benchmarking}
\input{7_conclusions}

\newpage
\bibliographystyle{IEEEtran}
\bibliography{refs}

\end{document}

%% file: 1_introduction.tex
\section{Introduction}

%%%%%%%%%%%%%%%% FIGURE %%%%%%%%%%%%%%%%%%%%
\begin{figure}[tp]
    \centering
    \includegraphics[width=0.8\linewidth]{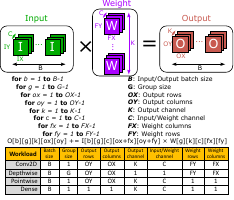}
    \caption{8-nested loop DNN layer representation and workloads representation}
    \label{fig:workload}
\end{figure}
%%%%%%%%%%%%%%%% FIGURE %%%%%%%%%%%%%%%%%%%%
%%%%%%%%%%%%%%%% FIGURE %%%%%%%%%%%%%%%%%%%%
\begin{figure*}[tbp]
    \centering
    \includegraphics[width=\linewidth]{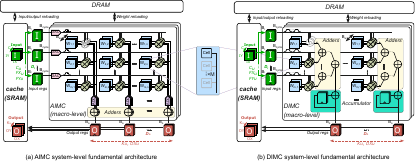}
    \caption{Fundamental architecture for AIMC/DIMC and their mapping paradigm. "u" in the subscript represents spatial unrolling.
    }
    \label{fig:imc_arch}
\end{figure*}
%%%%%%%%%%%%%%%% FIGURE %%%%%%%%%%%%%%%%%%%%

Recent developments of ultra-low power machine learning models have enabled the deployment of artificial intelligence on extreme edge devices.
However typical embedded digital accelerators suffer from high data movement costs and low computational densities, degrading the energy efficiencies by up to 2$\times$-1000$\times$ with respect to the ideal baseline of digital computations.
To minimize the data transfer overhead, in-memory computing (IMC) has recently emerged as a promising alternative to conventional accelerators based on arrays of digital processing elements (PEs). By directly performing the operations near/in the memory cells, these architectures allow to greatly reduce access overheads and enable massive parallelization opportunities, with potential orders of magnitude improvements in energy efficiency and throughput \cite{murmann_aimc}.
Most of the initial IMC designs published in the literature are focused on analog IMC (AIMC), where the computation is carried out in the analog domain \cite{aimc1,aimc2,aimc3}. While this approach ensures extreme energy efficiencies and massive parallelization, the analog nature of the computation and the presence of intrinsic circuit noise and mismatches compromises the output accuracy. 
To avoid the hurdles of AIMC, digital in memory computing (DIMC) is lately gaining more interest as a valid alternative \cite{dimc1,dimc2,dimc3,dimc4}. Thanks to its noise-free computation and more flexible spatial mapping possibilities, DIMC allows to realize additional flexibility and accurate computation at the expense of reduced peak energy efficiency.
These new opportunities stemming from AIMC and DIMC resulted in many recent implementations and publications in the literature. However, these works vary strongly in terms of hardware architecture, array dimensions, and silicon technology. This makes it difficult to grasp their relative strengths.
While several works assess and discuss IMC trends qualitatively, only few aim at quantitatively modeling or benchmarking architectural strategies.
This paper tries to fill this gap by: 
\begin{itemize}
    \item providing a unified analytical performance model for AIMC and DIMC, which is validated against several design points from the literature.
    \item performing quantitative performance benchmarking for various configurations
    of AIMC and DIMC, in terms of macro-level peak performance as well as system-level workload performance, by integrating the proposed model into ZigZag \cite{zigzag}, a system-level hardware design space exploration (DSE) framework.
\end{itemize}

%% file: 2_background.tex
\section{Background and related works}

\subsection{Dataflow concepts for IMC}

Deep neural network (DNN) workloads consist of a sequence of layers. The operations in the most common layers can be described as a combination of 8-nested for loops, which iterate over the indices of an input feature map tensor $I$, a weight tensor $W$ and that generate an output tensor $O$, as shown in Fig. \ref{fig:workload}.  
The tensor operations can be decomposed in a sequence of matrix-vector multiplications (MVM) by tiling the suitable loops.
These MVM operations offer a great opportunity for in-memory acceleration, as their dense 2D array structure aligns well with the array structure of the memory macros. 
Due to the physical properties of the hardware template, the spatial unrolling dimensions of the weight matrix are chosen so as to maximize spatial reuse of the activations along the columns of the memory, and accumulation of the partial sums along the rows of the IMC array, as shown in Fig. \ref{fig:imc_arch}: The $K$ and $OX$ loops -- (partially) irrelevant for the inputs -- are typically unrolled across the columns, while the $C$, $FX$ and $FY$ loops -- irrelevant for the outputs -- are parallelized across the rows. To further speed up computation and increase the reuse of operands at chip-level, the $OX$, $OY$ or $G$ loops can be parallelized across multiple IMC macros on the same die, requiring, however, duplication of the weights \cite{zigzag}.
It is nevertheless necessary to carefully align
the array dimensions and the workload parallelism to minimize underutilization of the computational resources.

\subsection{SRAM-based IMC}

SRAM is largely deployed for IMC, thanks to its robustness, large scale integration in CMOS, high endurance, and reliability, when compared to NVM-based solutions. 
SRAM-based IMC designs can be categorized into AIMC and DIMC based on whether the MAC computation is carried out in the analog or digital domain respectively.

\subsubsection{AIMC architectures}

{The AIMC architecture template is depicted in Fig.} \ref{fig:imc_arch}.a. 
To minimize data transfers between the compute array and the external memory levels and to maximize the array utilization, a weight stationary dataflow is commonly adopted for IMC designs. This dataflow aims at minimizing the data movement related to weights, maximizing their reuse at the computational array level: in IMC architectures this translates in pre-loading $B_w$-bit weights into the array, where 
$M$ memory cells are grouped and attached on the local bitline to increase memory density.
To execute an MVM, a digital input vector is provided along the wordlines of the memory array; the values of each element of the vector are converted to the analog domain with DACs -- either as a pulse width \cite{aimc1} or a voltage level \cite{aimc2, aimc3} -- and the resulting signals propagated along all wordlines in parallel. The analog signal on the wordlines is then combined with the value stored in the activated SRAM cells performing one multiplication per cell. The result of each multiplication will be transmitted onto the bitlines, where accumulation occurs in the analog domain across all cells connected to the same bitline. The final value on the bitlines is then converted back to the digital domain by means of ADCs, stored in output registers, after which it flows back to the higher level memories.

\subsubsection{DIMC architectures} 
In contrast to AIMC, the multiplication and accumulation operation in DIMC \cite{dimc1,dimc2,dimc3,dimc4} are implemented with digital logic gate-based multiplier and adder circuitry, as shown in Fig. \ref{fig:imc_arch}.b. IMC multiplication is done at SRAM cell level, where $M$ memory cells share one multiplying NAND gate.
The digital MAC results can immediately be offloaded to output registers after accumulating the full-precision results in digital accumulators. 
The presence of extra logic at cell level and the adder trees in DIMC give rise to area overheads and lower peak energy efficiencies compared to its analog counterpart;
yet, increased dataflow flexibility can be attained: the spatial loop unrolling at macro level can be made reconfigurable  by means of multiplexers in the accumulation trees with low energy and area overheads. 
A further benefit of DIMC lies in its robustness against noise and variability, enabling high operand precision. Thanks to its purely digital signal processing, DIMC computes deterministically, while the energy and area cost of DIMC scale proportionally with the operand precision. However, in literature the trade-off between efficiency and increased flexibility is not quantified on actual workloads and a back-to-back comparison of these two IMC paradigms is necessary to identify the optimal design choice. 

\begin{figure}[tb]
    \centering
    \includegraphics[width=\linewidth]{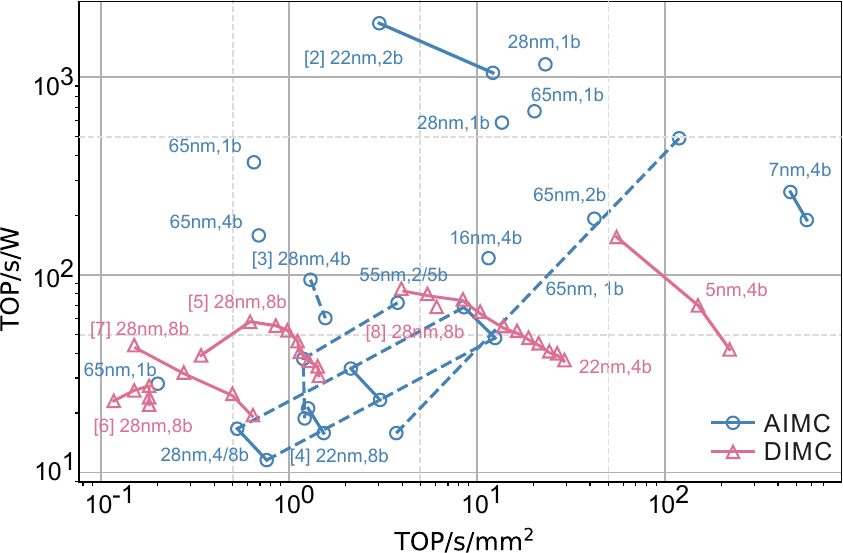}
    \caption{Peak performance benchmarking on AIMC/DIMC architectures. Each point reports the used technology node and weight bit precision.}
    \label{fig:benchmarking_fig_1}
\end{figure}

\subsection{IMC design benchmarking}
Numerous implementations for AIMC and DIMC exist in literature, however they differ in their hardware architectures, array dimensions, and fabrication technologies, making it challenging to grasp their relative strengths \cite{imc_trends_1, imc_trends_2, burr_benchmarking, murmann_aimc, imc_trends_3, imc_trends_4}.
More importantly, a fundamental concern in the reported performance metrics is that IMC chips are mostly benchmarked according to their peak performance for a single IMC macro, as summarized in Fig. \ref{fig:benchmarking_fig_1}. However, these performance assessments unrealistically suppose that the (in-memory) computing logic is fully utilized at all times: to evaluate the performances on actual workloads, DSE frameworks and cost models have been implemented to explore the design space and design the hardware based on the target network model.

\subsection{DSE framework and modeling tools}

In order to evaluate the two IMC design paradigms at system level on target workloads design space exploration at the system level can be done through analytical models for IMC. Published models however primarily focus on AIMC designs \cite{9218657, imc_modeling_2, murmann_aimc, imc_modeling_3, imc_modeling_4, imc_modeling_5, imc_modeling_6, imc_modeling_8}, while there is a lack of DIMC modeling efforts. Similarly, for mapping space explorations, most of the focus has been dedicated to AIMC designs, while lacking DIMC assessment \cite{imc_dse_1, imc_modeling_4, imc_modeling_5, imc_dse_2}. In particular, the work in \cite{9218657} offers analytical energy and area modeling of AIMC ADCs and DACs, but the model is only validated on standalone ADCs and DACs rather than actual IMC implementations.
An alternative energy model is proposed in \cite{imc_modeling_2}, however, the energy cost of each component is separately extracted from different papers, and has not been validated against a complete taped-out chip.

Fully digital ML processors, in contrast, have been the subject of many system-level design space explorations:
dedicated DSE frameworks supporting hardware-mapping co-optimization have emerged over the last years \cite{zigzag, timeloop, maestro}, with different levels of supported parametrization of the hardware architecture space.
Yet, so far none of these frameworks have been deployed to extensively assess AIMC as well as DIMC topologies, due to a lack of dedicated cost modeling efforts. To address this gap, we present an analytical IMC cost model in the following section.

%% file: 4_imc_model.tex
\section{Unified analytical IMC performance model}
\label{sec:modeling}

\input{table_model_summary}
\input{table_constants}

The fundamental structure of an IMC accelerator is constant across different implementations: a computational memory array with an interface to read and write data from. However, large variability exists in the implementation details of each work. 
While it is a daunting task to develop an accurate unified analytical cost model that matches all designs from literature, it is possible to create a model within an acceptable mismatch for system-level evaluation, enabling us to identify the main bottlenecks and the benefits of various architectural choices.

To describe the granularity of the computational parallelization options in the IMC array, the presented model represents the array along a set of dimensions, as depicted in Fig. \ref{fig:imc_arch}: $D_i$ refers to the parallelism of the input activation vector, while $D_o$ refers to the parallelism of the output vector generated after each MVM operation. Each element along $D_o$ is further split into separate cells based on the weight bitwidth $B_w$; similarly, each element along the $D_i$ dimension corresponds to $M$ cells, both in AIMC \cite{aimc2, 9731681, 10067526} and DIMC \cite{dimc2, dimc3, dimc4}. Moreover, the MVM operation can be temporally split in multiple cycles, in which $B_{\text{cycle}}$ bits are processed per cycle per input.

To estimate area, energy and delay for both DIMC and AIMC macros, a unified analytical cost model is developed for these architectures, to enable performance evaluation at the system level for different design points. The model takes into account memory instances, analog peripherals, and digital peripherals for the energy, area and latency estimation, while the cost of each component is shown in Table \ref{tab:table1}, together with its calibration parameters for 28nm.
The cost models for AIMC and DIMC for each of these components will be further detailed in the next subsections. 

%%%%%%%%%%%%%%%% FIGURE %%%%%%%%%%%%%%%%%%%%
\begin{figure}[tb]
    \centering
    \includegraphics[width=0.6\linewidth]{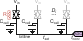}
    \caption{Equivalent worst case scenario for ADC in AIMC macro. The transistor or transmission gate in each IMC cell connected to the bitline can be treated as a resistor and a capacitor. The worst case happens when only one cell is charging or discharging all capacitors.}
    \label{fig:adc_model}
\end{figure}
%%%%%%%%%%%%%%%% FIGURE %%%%%%%%%%%%%%%%%%%%

\subsection{Memory instances: IMC cell array and registers}

Memory instances refer to the set of storage elements in the IMC macro: these are the weight memory cells in the computational array, as well as the input and output registers.
Assuming that the weight memory elements in the macro are based on 6T SRAM cells, CACTI \cite{cacti} is utilized to estimate the area, delay and energy per access of the IMC cell array. The CACTI model, based on an RC model, is further validated with synthesized memory macros via a memory compiler. 
However the CACTI model does not correctly cover the MVM computation in IMCs since multiple wordlines activate simultaneously for AIMC and there is no significant bitline activity for DIMC during the computation, behaviours that do not correspond to normal SRAM operation. 
Given this, a separate model is adopted for IMC MVM array computation.
For AIMC, assuming that energy is mostly consumed in the charge and discharge phases of the bitlines, the total energy cost on bitline switching (E$_{\text{cell array}}$) is expressed as:

\begin{equation}
    E_{\text{cell array}} = C_{\text{cell}}  V_{\text{DD}}^2  B_w \cdot D_i \cdot D_o
    \label{eq:e_bl}
\end{equation}
Two transistors seen by the bitline for each bitcell, $C_{\text{cell}}$ is normalized to 0.5$C_{\text{gate}}$, where $C_{\text{gate}}$ represents the gate capacitance of a four-transistor NAND2/NOR2 gate. In DIMC, with its very short bitlines, this cost is negligible.

%%%%%%%%%%% figure %%%%%%%%%%%%%%
\begin{figure}[tb]
    \centering
    \includegraphics[width=\linewidth]{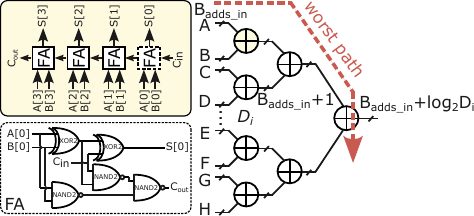}
    \caption{$B_{\text{adds\_in}}$-bit adder tree topology for A/DIMC (right). $B_{\text{adds\_in}}=\text{ADC}_{\text{res}}$ for AIMC and $B_{\text{adds\_in}}=B_w$ for bit-serial DIMC. The structure of each node (single-stage adder) assuming RCA type is shown in the top left. The gate-level circuit for 1b FA is shown in the bottom left.
    }
    \label{fig:rca_tree}
\end{figure}
%%%%%%%%%%% figure %%%%%%%%%%%%%%

The delay of the signal traversal on the bitline is merged into the ADC delay for AIMC. As DIMC has negligible delay on the local bitline, the delay cost attached to the cell array can be approximated to 0 for the MAC computation. The energy and area cost of input and output registers are modeled as D Flip-Flop (DFF) standard cells, using the relative cost ratio of a DFF to a NAND2 gate (Table \ref{tab:table1}). 
The clock-to-Q delay $D_{\text{DFF}}$ is negligible compared to other peripheral contributions.

\subsection{Analog peripherals: ADCs and DACs}

AIMC requires DACs and ADCs to interface with the computational array to initiate/terminate the array operation. Their contributions are described in this section.

\subsubsection{ADC model} The largest cost in AIMC designs are usually associated with ADCs and is strongly dependent on the ADC resolution.
Following \cite{murmann_aimc} and assuming uncorrelated and uniformly distributed activation and weights, the required ADC resolution is determined by the activation precision $B_i$ and the amount of accumulation terms $D_i$: 

\begin{equation}
    \text{ADC}_{\text{res}} = B_i + \log_2(k \cdot \text{FS} \cdot \sqrt{D_{i}})
    \label{eq:adc_res}
\end{equation}
with FS$=0.5$ (as in \cite{murmann_aimc}) representing the ADC's normalized full-scale range occupying 50\% of the entire voltage range; the empirical constant $k=2$ represents the ADC noise deviation at the output to be $k$ times smaller than the accumulation error variance on the bitline. Using the resolution of Eq. (\ref{eq:adc_res}) an energy, latency, and area model is developed assuming a SAR-ADC topology, the most common ADC topology deployed for AIMC designs thanks to its small area footprint and low energy consumption.
The energy model, based on \cite{murmann_aimc}, is defined as:
\begin{equation}
    E_{\text{ADC}} = (k_1  \text{ADC}_{\text{res}} + k_2  4^{\text{ADC}_{\text{res}}}) V_{\text{DD}}^2
    \label{eq:eq_Eadc}
\end{equation}
with constants $k_1$ and $k_2$ obtained by fitting to the most energy efficient ADC designs, dominated by SAR-ADCs \cite{murmann_survey}. The ADC delay cost is expressed as the sum of 1) the settling time of the analog value on the bitline $t_{\text{set}}$ and 2) the time required for the ADC conversion $t_{\text{conv}}$. The first contribution can be described as the delay of a parallel RC network at the ADC input node (Fig. \ref{fig:adc_model}). 
The time required for the signal on the bitline to settle within 1 LSB of the ADC is given in Eq. (\ref{eq:adc1}), with the RC time constant $\tau=D_i \cdot R_{\text{cell}}C_{\text{cell}}$. The model assumes the worst case scenario for $t_{\text{set}}$, where only one cell is active on the entire bitline.

\begin{equation}
\begin{split}
    & V_{\text{DD}} (1-e^{-t_{\text{set}}/\tau}) < V_{\text{DD}} (1/2^{\text{ADC}_{\text{res}}}) \\
    & \Rightarrow t_{\text{set}} > 0.69\tau \text{ADC}_{\text{res}}
    \label{eq:adc1}
\end{split}
\end{equation}

The second contribution, the time required for the ADC conversion, is defined as in Eq. (\ref{eq:adc2}), where the delay for a single bit conversion $t_{\text{conv/bit}}$ is a constant that depends on the technology node.
\begin{equation}
    t_{\text{conv}} = t_{\text{conv/bit}} \cdot \text{ADC}_{\text{res}}
    \label{eq:adc2}
\end{equation}

By combining Eq. (\ref{eq:adc1}) and Eq. (\ref{eq:adc2}), the ADC delay can be modeled in Eq. (\ref{eq:adc3}), where $k_3 = 0.69 R_{\text{cell}}C_{\text{cell}}$ reflects the impact of the number of cells attached to the bitline and $k_4=t_{\text{conv/bit}}$ represents the intrinsic conversion delay per bit for SAR ADC. Table I lists the used parameters $k$, fitted for 28nm technology.

\begin{equation}
    D_{\text{ADC}} = (k_3 \cdot D_i+k_4) \cdot \text{ADC}_{\text{res}}\
    \label{eq:adc3}
\end{equation}

Based on \cite{9218657}, the ADC area is estimated as:

\begin{equation}
    A_{\text{ADC}} = 10^{-k_5  \text{ADC}_{\text{res}} + k_6} \cdot 2^{\text{ADC}_{\text{res}}}
    \label{eq:eq_Aadc}
\end{equation}

\subsubsection{DAC model} The delay and area costs of the DACs are neglected, as they have relatively small latency overheads and area \cite{aimc1}. The DAC energy cost is modeled using Eq. (\ref{eq:eq_Edac}), where $k_7$ is a technology constant that represents the capacitance associated with each bit conversion.

\begin{equation}
    E_{\text{DAC}} = k_7 \text{DAC}_{\text{res}} V_{\text{DD}}^2\ 
    \label{eq:eq_Edac}
\end{equation}

\subsection{Digital peripherals}
 
\subsubsection{Digital multiplication} In both AIMC and DIMC, the multiplication operation is modeled
at the cell level as (one or more) 1-bit operation(s) using a NAND gate between the 1-bit input and 1-bit weight stored in the cell. Hence, the cost of such 1-bit multiplier is equivalent to the cost of a single gate, as listed in Table \ref{tab:table1}.

\subsubsection{Digital accumulation} Both AIMC and DIMC use adder trees to combine digital outputs and get eventually partial sums across separate cycles.
In the proposed model, these partial sum adder trees and accumulators are modeled assuming ripple-carry adders (RCA) (Fig. \ref{fig:rca_tree}). The input precision $B_{\text{adds\_in}}$ equals weight resolution $B_w$, $\text{ADC}_{\text{res}}$ for DIMC, AIMC respectively (Fig. \ref{fig:imc_arch}).
The cost of the adder trees is expressed as a function of its basic unit, the Full Adder (FA). The circuit of a standard FA is shown in Fig. \ref{fig:rca_tree}, consisting of three NAND2 gates and two XOR2 gates, with its cost computed as listed in Table \ref{tab:table1}. 
The total energy and area of an adder tree depends on the number of FA, which can be expressed as below:

\begin{equation}
    \#\text{FA} = \sum^{\log_2D_i}_{n=1} (B_{adds\_in} + n - 1) \frac{D_i}{2^n}
    \label{eq:eq_nbfa}
\end{equation}

The delay of the adder tree $D_{\text{adds}}$ depends on the number of FA on the critical path, as shown in Fig. \ref{fig:rca_tree}. This component can be modeled as in Eq. (\ref{eq:eq_Dfa_sum}), where the output precision of the adder tree $B_{\text{adds\_out}}=B_{\text{adds\_in}}+\log_2D_i$.

\begin{equation}
    D_{\text{adds}} = D_{\text{FA}_S}\log_2D_i + D_{\text{FA}_C}B_{\text{adds\_out}}
    \label{eq:eq_Dfa_sum}
\end{equation}

Finally, the cost of the accumulator can be derived from the cost of the FA and a group of DFFs. The costs of the RCA-based adder trees and accumulators for both AIMC and DIMC macros are all summarized in Table \ref{tab:table1}.

%% file: table_model_summary.tex
\begin{table*}[tb]
\centering
\setlength{\extrarowheight}{3pt}

\caption{Cost summary for components in a single IMC macro}

\begin{tabularx}{\linewidth}{|X|X|XX|XXX|}
\hline

% row #1
\multicolumn{1}{|l|}{\textbf{Component}} &
  \multicolumn{2}{c|}{\textbf{Energy}} &
  \multicolumn{2}{c|}{\textbf{Delay}} &
  \multicolumn{2}{c|}{\textbf{Area (\SI{}{\micro\meter}$^2$)}} \\ \cline{2-7} 

% row #2
\multicolumn{1}{|l|}{} &
  \multicolumn{1}{c|}{\textbf{AIMC}} &
  \multicolumn{1}{c|}{\textbf{DIMC}} &
  \multicolumn{1}{c|}{\textbf{AIMC}} &
  \multicolumn{1}{c|}{\textbf{DIMC}} &
  \multicolumn{1}{c|}{\textbf{AIMC}} &
  \multicolumn{1}{c|}{\textbf{DIMC}}  \\ \hline

% row #3 (IMC cell array)

  \textbf{Cell array} &
  \multicolumn{1}{c|}{$0.5C_{\text{gate}}V_{\text{DD}}^2B_w \cdot D_i \cdot D_o$} &
  \multicolumn{1}{c|}{0} &
  \multicolumn{2}{c|}{0} &
  \multicolumn{2}{c|}{CACTI \cite{cacti}} \\ \hline

% row #4 (ADC)

  \textbf{ADC} &
  \multicolumn{1}{l|}{$(k_1\text{ADC}_{\text{res}}+k_24^{\text{ADC}_{\text{res}}})V_{\text{DD}}^2$} &
  \multicolumn{1}{c|}{--} &
  \multicolumn{1}{l|}{$(k_3 \cdot D_i + k_4)\text{ADC}_{\text{res}}$} &
  \multicolumn{1}{c|}{--} &
  \multicolumn{1}{l|}{$10^{-k_5 \text{ADC}_{\text{res}} + k_6} \cdot 2^{\text{ADC}_{\text{res}}}$} &
  \multicolumn{1}{c|}{--} \\ \cline{1-7} 

% row #5 (DAC)

  \textbf{DAC} &
  \multicolumn{1}{c|}{$k_7 \cdot \text{DAC}_{\text{res}}V_{\text{DD}}^2$} &
  \multicolumn{1}{c|}{--} &
  \multicolumn{1}{c|}{0} &
  \multicolumn{1}{c|}{--} &
  \multicolumn{1}{c|}{0} &
  \multicolumn{1}{c|}{--}  \\ \hline

% row #6 (Multiplier)

  \textbf{1b multiplier} &
  \multicolumn{2}{c|}{$0.5C_{\text{gate}}V_{\text{DD}}^2$} &
  \multicolumn{2}{c|}{$D_{\text{gate}}$} &
  \multicolumn{2}{c|}{$A_{\text{gate}}$} \\ \cline{1-7} 

% row #7 (adder tree)

  \textbf{Adder tree} &
  \multicolumn{2}{c|}{$E_{\text{FA}} \cdot \#\text{FA}$} &
  \multicolumn{2}{c|}{$D_{\text{FA}_S}log_2D_i + D_{\text{FA}_C}B_{\text{adds\_out}}$} &
  \multicolumn{2}{c|}{$A_{\text{FA}} \cdot \#FA$}
   \\ \cline{1-7} 

% row #7 (accumulator)

  \textbf{Accumulator} &
  \multicolumn{2}{c|}{$(E_{\text{FA}}+E_{\text{DFF}})B_{\text{acc}}$} &
  \multicolumn{2}{c|}{$D_{\text{FA}_C}(B_{\text{acc}}-B_{\text{adds\_out}})$} &
  \multicolumn{2}{c|}{$(A_{\text{FA}}+A_{\text{DFF}})B_{\text{acc}}$} 
   \\ \hline

% row #8 (input/output registers)

  \textbf{1b register} &
  \multicolumn{2}{c|}{$E_{\text{DFF}}$} &
  \multicolumn{2}{c|}{$0$} &
  \multicolumn{2}{c|}{$A_{\text{DFF}}$} 
   \\ \hline
\end{tabularx}

% add note
\begin{tabularx}{\linewidth}{XX}
    $C_{\text{gate}}, D_{\text{gate}}, A_{\text{gate}}$: capacitance, delay, area of single NAND2 gate & 
    %$V_{DD}$: supply voltage (0.9 V @ 28nm) \\
    %(0.7 fF, 47.8 ps, 0.614 $\mu m^2$ @ 28nm) & \\
    $B_i, B_w, B_o$: input, weight, output precision in bit \\
    $V_{\text{DD}}, D_i, D_o$: supply voltage, size of input dimension, output dimension &
    % R_{cell}: 27 k$\Omega$
    %$k_1, k_2, k_3, k_4, k_5, k_6, k_7$: 100 fF, 1 aF, 6.53 ps, 640 ps, 0.0369, 1.206, 50 fF @ 28nm &
    $\text{ADC}_{\text{res}}, \text{DAC}_{\text{res}}$: ADC resolution, DAC resolution \\
    $B_{\text{adds\_in}}$, $B_{\text{adds\_out}}$, $B_{\text{acc}}$: input precision of an adder tree, output precision of an adder tree, output precision of an accumulator &
    $E_{\text{FA}}, D_{\text{FA}_S}, D_{\text{FA}_C}, A_{\text{FA}}, \#$FA: energy/sum-delay/carry-delay/area of FA, number of FA in an adder tree
    %$E_{DFF}, A_{DFF}$: energy and area of DFF ($3C_{gate}  V_{DD}^2$, $6A_{gate}$ @ 28nm) \\
    %\multicolumn{2}{l}{$E_{FA}, D_{FA\_S}, D_{FA\_C}, A_{FA}$: energy/sum-delay/carry-delay/area of FA} ($6C_{gate}  V_{DD}^2$, $4.8D_{gate}$, $2D_{gate}$, $7.8A_{gate}$ @ 28nm)} \\
    %\multicolumn{2}{l}{$\#FA$: number of FA in an adder tree. For a RCA tree with $D_i$ $B_{in}$-b inputs, $\#FA=\sum^{\log_2D_i}_{n=1} (B_{in} + n - 1) \frac{D_i}{2^n}$, $B_{adds}=B_{in}+log_2D_i$}
\end{tabularx}
\label{tab:table1}
\end{table*}

%% file: table_constants.tex
\begin{table}[]
\centering

\caption{Parameters in proposed cost model @ 0.9V, 28nm}
\begin{tabular}{|l|l|l|l|}

\hline
\textbf{Parameter} & \textbf{Value} & \textbf{Parameter} & \textbf{Value}               \\ \hline
$C_{\text{gate}}$        & 0.7 fF                     & k$_6$              & 1.206                                    \\ %\hline
$D_{\text{gate}}$        & 47.8 ps                    & k$_7$               & 50 fF                                    \\ %\hline
$A_{\text{gate}}$        & \SI{0.614}{\micro\meter}$^2$              & $E_{\text{FA}}$          & 6$C_{\text{gate}}V_{\text{DD}}^2$ \\ %\hline
k$_1$               & 100 fF \cite{murmann_aimc}                    & $D_{\text{FA}_S}$       & 4.8$D_{\text{gate}}$                           \\ %\hline
k$_2$               & 1 aF \cite{murmann_aimc}                      & $D_{\text{FA}_C}$       & 2$D_{\text{gate}}$                             \\ %\hline
k$_3$               & 6.53 ps                    & $A_{\text{FA}}$          & 7.8$A_{\text{gate}}$                           \\ %\hline
k$_4$               & 640 ps                     & $E_{\text{DFF}}$         & 3$C_{\text{gate}}V_{\text{DD}}^2$ \\ %\hline
k$_5$               & 0.0369                     & $A_{\text{DFF}}$         & 6$A_{\text{gate}}$                             \\ \hline
\end{tabular}
\end{table}

%% file: 5_model_validation.tex
\section{Macro-level model validation}

\begin{figure*}[tbp]
    \centering
    \includegraphics[width=\textwidth]{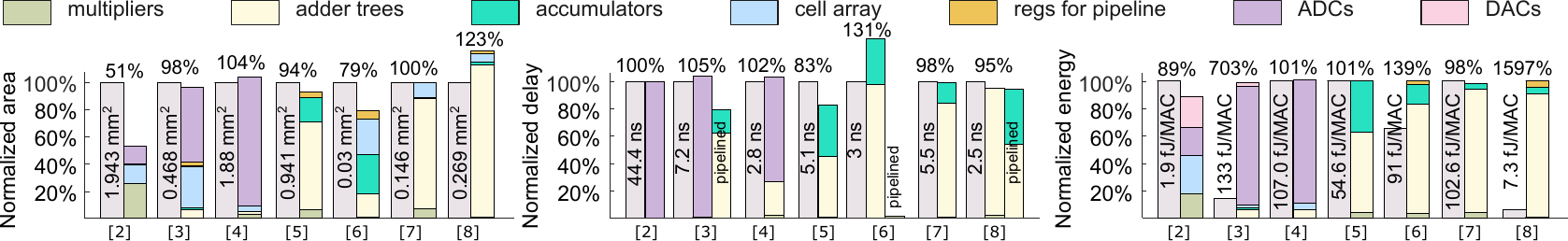}
    \begin{tabularx}{\linewidth}{XX}
    \footnotesize
    - \cite{aimc1}: reported area includes repeaters, decap, et.al. \\
    \footnotesize
    - \cite{aimc1, aimc3}: area and energy estimation has been linearly scaled from 28nm to 22nm.  \\
    \footnotesize
    - \cite{aimc2, dimc2, dimc4}: pipeline inserted before or in the middle of the adder tree, which decreases the delay by 0.5$\times$.
    \\
    \footnotesize	
    - 50\% input toggle rate and 50\% weight sparsity are assumed in model validation. These values are clearly reported in \cite{aimc1,aimc3,dimc1,dimc3}, yet sparsity is not reported in other papers.
    \end{tabularx}
    \caption{Macro-level validation of proposed IMC cost model against 28nm/22nm implementations. Gray bars: published metrics; Colored bar: model estimation.
    }
    \label{fig:model_validation}
\end{figure*}

The model presented in Section~\ref{sec:modeling}  is validated against the area, delay, and energy consumption reported of a collection of taped-out IMC chips for both AIMC and DIMC architectures (Fig. \ref{fig:model_validation}). To decouple the impact of the chosen technology node from the performance and efficiency metrics, only published IMC implementations in 22nm and 28nm technology, have been considered.
Table \ref{tab:table2} lists the considered implementations, together with their hardware configuration. 
The following paragraphs will analyze the validation results in more detail.

%%%%%%%%%%% figure %%%%%%%%%%%%%%
%%%%%%%%%%% figure %%%%%%%%%%%%%%

\subsection{AIMC macro-level model validation}

The validation of the model against 22-28nm AIMC designs was carried out on \cite{aimc1, aimc2, aimc3}.
The IMC arrays of these works all consist of 6T SRAM cells followed by SAR ADCs.

For the area validation the model shows a good match, except for \cite{aimc1}, which extensively uses repeaters in their memory banks. This results in a significantly larger(4.5$\times$) SRAM array compared to the CACTI estimation.
With the extracted Eq. (\ref{eq:adc3}), the delay validations match well. And finally,
for the energy estimation a maximum mismatch of 11\% is achieved, except for \cite{aimc2}, which is believed to be caused by difference in sparsity used in model and in their chip measurements. Unfortunately, the used sparsity is not reported in \cite{aimc2}.

\input{table_validation_papers}

\subsection{DIMC macro-level model validation}

The validation of the model against DIMC designs was carried out on \cite{dimc1, dimc2, dimc3, dimc4}, which are all implemented in 28nm technology with 6T SRAM cells and all employ the digital addition operation with conventional RCA structures.
Although different configurations are adopted in these implementations (i.e., different $M$ and $\#$macros), the model shows a good area estimation for \cite{dimc1} and \cite{dimc3}, while it underestimates \cite{dimc2}, which contains additional logic to support adder tree configurability, and overestimate \cite{dimc4}, which introduces approximate multipliers for increased efficiency. In terms of delay, the model matches well, with some mismatch for \cite{dimc1}, due to delay in its unmodeled Booth Encoder, and for \cite{dimc2}, due to its configurable adder tree topology.
For the energy validation, the model estimation excellently fits the design of \cite{dimc1, dimc3} which clearly reported their sparsity in the publication. \cite{dimc2,dimc4} did not report the used sparsity in their measurements. Our 50\% sparsity assumption seems to be significantly off from their unreported measurement settings.

Although mismatches can be observed whenever special approximate computing or extreme sparsity techniques are used, the mismatch is limited to 20\% for standard implementations. It can therefore serve as an excellent vehicle to assess the impact of various architectural parameters on system-level performance under peak/actual workloads in the next section.

%% file: table_validation_papers.tex
\begin{table}[tbp]
\centering
\setlength{\extrarowheight}{3pt}

\caption{Hardware configurations of the validation works  %\\
%\mv{we miss Bcycle here?} \mv{Keep close to figure 6} \jsnote{fixed}
}
%\mv{[Add a column with "AIMC" or "DIMC"]}
\begin{tabular}{|c|c|c|c|c|c|c|}
\hline

% row #1 (column title)
\multicolumn{1}{|l|}{\textbf{Index}} &
  \multicolumn{1}{c|}{\textbf{IMC}} &
  \multicolumn{1}{c|}{\textbf{$\boldsymbol{B_i/B_w/B_{cycle}}$}} &
  %\multicolumn{1}{c|}{\textbf{$B_{cycle}$}} &
  \multicolumn{1}{c|}{\textbf{$\boldsymbol{D_i}$}} &
  \multicolumn{1}{c|}{\textbf{$\boldsymbol{D_o}$}} &
  \multicolumn{1}{c|}{\textbf{$\boldsymbol{M}$}} &
  \multicolumn{1}{c|}{\textbf{$\boldsymbol{\#\text{macros}}$}} \\   \hline

\multicolumn{1}{|l|}{\text{\cite{aimc1}}} &
  \multicolumn{1}{c|}{\text{AIMC}} &
  \multicolumn{1}{c|}{\text{7/2/7}} &
  %\multicolumn{1}{c|}{\text{7}} &
  \multicolumn{1}{c|}{\text{1024}} &
  \multicolumn{1}{c|}{\text{512}} &
  \multicolumn{1}{c|}{\text{1}} &
  \multicolumn{1}{c|}{\text{1}} \\   \hline

\multicolumn{1}{|l|}{\text{\cite{aimc2}}} &
  \multicolumn{1}{c|}{\text{AIMC}} &
  \multicolumn{1}{c|}{\text{8/8/2}} &
  %\multicolumn{1}{c|}{\text{2}} &
  \multicolumn{1}{c|}{\text{16}} &
  \multicolumn{1}{c|}{\text{12}} &
  \multicolumn{1}{c|}{\text{32}} &
  \multicolumn{1}{c|}{\text{1}} \\   \hline
  
\multicolumn{1}{|l|}{\text{\cite{aimc3}}} &
  \multicolumn{1}{c|}{\text{AIMC}} &
  \multicolumn{1}{c|}{\text{8/8/1}} &
  %\multicolumn{1}{c|}{\text{1}} &
  \multicolumn{1}{c|}{\text{64}} &
  \multicolumn{1}{c|}{\text{256}} &
  \multicolumn{1}{c|}{\text{1}} &
  \multicolumn{1}{c|}{\text{8}} \\   \hline

\multicolumn{1}{|l|}{\text{\cite{dimc1}}} &
  \multicolumn{1}{c|}{\text{DIMC}} &
  \multicolumn{1}{c|}{\text{8/8/2}} &
  %\multicolumn{1}{c|}{\text{2}} &
  \multicolumn{1}{c|}{\text{32}} &
  \multicolumn{1}{c|}{\text{6}} &
  \multicolumn{1}{c|}{\text{1}} &
  \multicolumn{1}{c|}{\text{64}} \\   \hline

\multicolumn{1}{|l|}{\text{\cite{dimc2}}} &
  \multicolumn{1}{c|}{\text{DIMC}} &
  \multicolumn{1}{c|}{\text{8/8/1}} &
  %\multicolumn{1}{c|}{\text{1}} &
  \multicolumn{1}{c|}{\text{32}} &
  \multicolumn{1}{c|}{\text{1}} &
  \multicolumn{1}{c|}{\text{16}} &
  \multicolumn{1}{c|}{\text{2}} \\   \hline

\multicolumn{1}{|l|}{\text{\cite{dimc3}}} &
  \multicolumn{1}{c|}{\text{DIMC}} &
  \multicolumn{1}{c|}{\text{8/8/2}} &
  %\multicolumn{1}{c|}{\text{2}} &
  \multicolumn{1}{c|}{\text{128}} &
  \multicolumn{1}{c|}{\text{8}} &
  \multicolumn{1}{c|}{\text{8}} &
  \multicolumn{1}{c|}{\text{8}} \\   \hline

\multicolumn{1}{|l|}{\text{\cite{dimc4}}} &
  \multicolumn{1}{c|}{\text{DIMC}} &
  \multicolumn{1}{c|}{\text{8/8/1}} &
  %\multicolumn{1}{c|}{\text{1}} &
  \multicolumn{1}{c|}{\text{128}} &
  \multicolumn{1}{c|}{\text{8}} &
  \multicolumn{1}{c|}{\text{2}} &
  \multicolumn{1}{c|}{\text{4}} \\   \hline

\end{tabular}
\label{tab:table2}
\end{table}

%% file: 6_benchmarking.tex
\section{System-level benchmarking for IMC}

The developed IMC analytical cost model can now be used to perform a set of architecture explorations to aid in design decisions. In this section, this will be pursued in the context of IMC architectures for edge devices. 
The focus will first be at macro level, where \textit{peak IMC efficiency} will be assessed, together with its scalability with array dimensions. This will be contrasted with \textit{peak system performance} -- i.e. also taking data transfers and activation memory area into account -- causing a shift in the optimal hardware configurations (Subsection V.A). In the second subsection, actual \textit{workload system performance} will be analyzed on MLPerf Tiny benchmark models \cite{mlperf}, studying how it differs from peak system performance across various workloads and the resulting optimal hardware configurations (Subsection V.B). For these system-level assessments, the derived cost models are integrated into a modified version of the system-level design space exploration tool ZigZag \cite{zigzag}. We perform this study based on a single IMC macro as shown in Fig.~\ref{fig:imc_arch}, with $M=1$ and assuming a 256 KB on-chip cache, which is equipped with a bandwidth fitting with IMC dimension and whose cost is estimated by CACTI \cite{cacti}. 
All assessments are carried out assuming INT8 operations, being the preferred quantization scheme in MLPerf Tiny models \cite{mlperf} with zero sparsity. 
$B_{\text{cycle}}$ is fixed to 2 and 1 for AIMC and DIMC, respectively, in light of their popularity in literature. 

\subsection{Peak performance assessment for A/DIMC designs}
Peak performance is the starting metric for evaluation as it provides a roofline for the computational capabilities of the design. In literature this is often the preferred way to quantify the efficiency of an IMC architecture, even though it does not provide a clear overview of the actual workload efficiency of the design \cite{TOPSW_harmful} (see Subsection V.B). All assessments in this section assume dense GeMM workloads filling the complete array with infinite weight reuse (weights remain always stationary).

\begin{figure}[tbp]
    \centering
    \includegraphics[width=\linewidth]{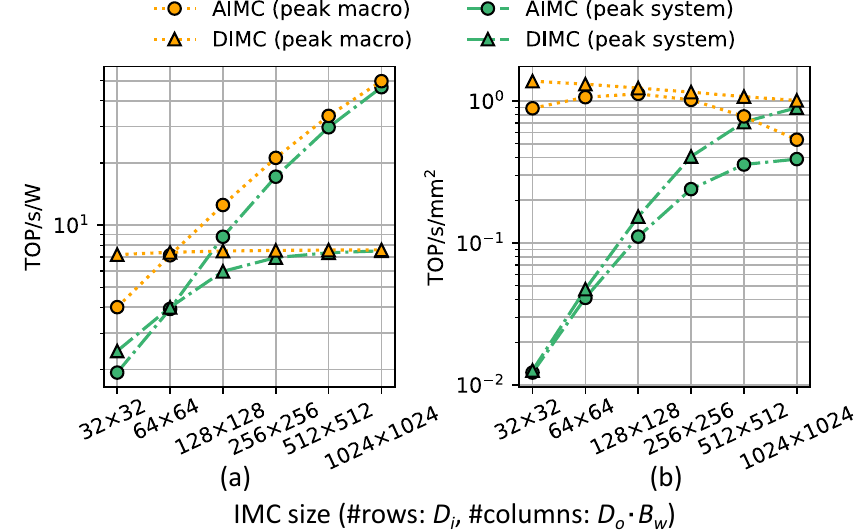}
    \caption{Peak macro performance and peak system performance with scaling IMC sizes
    }
    \label{fig:macro_vs_system}
\end{figure}

\begin{figure}[tbp]
    \centering
    \includegraphics[width=\linewidth]{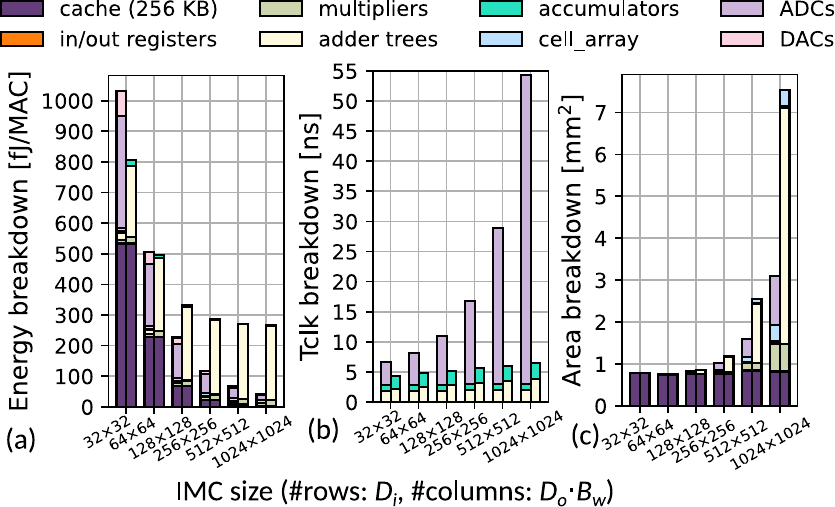}
    \caption{Peak system cost breakdown with scaling IMC sizes}
    \label{fig:area_delay_energy_peak}
\end{figure}

\paragraph{Macro efficiency} Fig.\ref{fig:macro_vs_system} shows the peak macro performance (excluding the cost of on-chip memories and off-chip DRAM) in function of IMC array size. The peak energy efficiency of AIMC is clearly depending on the amortization of its peripherals. 
An order of magnitude improvement is achieved by going from a 32$\times$32 array to a 1024$\times$1024 array, thanks to the reduction of the contribution of ADC/DAC conversions (Fig.\ref{fig:area_delay_energy_peak}.a). However, to preserve SNR at the analog output for these larger array sizes, ADCs with higher resolution are required, which in turn require longer conversion times and larger area overheads. This leads to degraded throughput and area, which affect the computational density in terms of TOP/s/mm$^2$ (Fig.\ref{fig:area_delay_energy_peak}.b).
DIMC energy efficiency, on the other hand, does not benefit from larger array sizes, since the cost of the adder trees and multipliers scale proportionally with the array size. While DIMC with a smaller size ($\leq 64 \times 64$) outperforms AIMC in terms of TOP/s/W this is no longer true for larger IMC arrays, in which the ADC overhead is better amortized.
The peak macro computational density for DIMC marginally decrease with array size due to the logarithmic increase of the depth of the adder trees. In general, DIMC consistently outperforms AIMC in terms of computational density, as the delay of the ADCs is worse than that of digital adder trees, enabling it to operate at a higher clock speed.
\paragraph{System efficiency} Fig.\ref{fig:area_delay_energy_peak} displays, next the the peak macro efficiency, also the peak system efficiency, including the cost of feeding data to/from on-chip feature memories and off-chip DRAM. For the 32$\times$32 scenario, the system energy efficiency is at least 2$\times$ below the macro efficiency. The gap between system and macro efficiency, however, diminishes when the array size increases thanks to the memory access cost amortization from multicasting the input and output operands across a larger array. 
Computational intensity (TOP/s/mm$^2$) is even more affected: for smaller IMC sizes, the cache area dominates the total system area and reduces the TOP/s/mm$^2$ by more than one order of magnitude. These trends can also clearly be observed from the energy, delay and area breakdowns at system level in Fig.~\ref{fig:area_delay_energy_peak}.

The takeaway might seem that larger arrays lead to overall better system efficiency. Yet, a massive array is useless if only a small portion of it is effectively utilized by actual workloads, since it leads to the nullification of any peripheral amortization. It is therefore necessary to study the efficiency under actual workload scenarios, as done in the next subsection.

\subsection{Actual workload performance for A/DIMC}

\begin{figure}[tbp]
    \centering
    \includegraphics[width=0.9\linewidth]{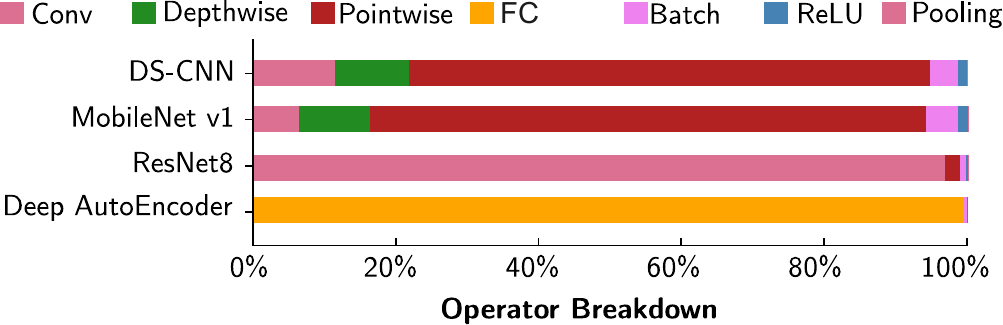}
    \caption{Operator breakdown of MLPerf Tiny \cite{mlperf} benchmark models 
    }
    \label{fig:mlperf_workload}
\end{figure}

Efficiency and performance metrics are estimated at system level on the MLPerf Tiny benchmark models, which exhibit a nice range of different layer topologies (Fig. \ref{fig:mlperf_workload}).

%%%%%%%%%%%%%%%% FIGURE %%%%%%%%%%%%%%%%%%%%
%%%%%%%%%%%%%%%% FIGURE %%%%%%%%%%%%%%%%%%%%

%%%%%%%%%%% figure %%%%%%%%%%%%%%
\begin{figure}[tbp]
    \centering
    \includegraphics[width=\linewidth]{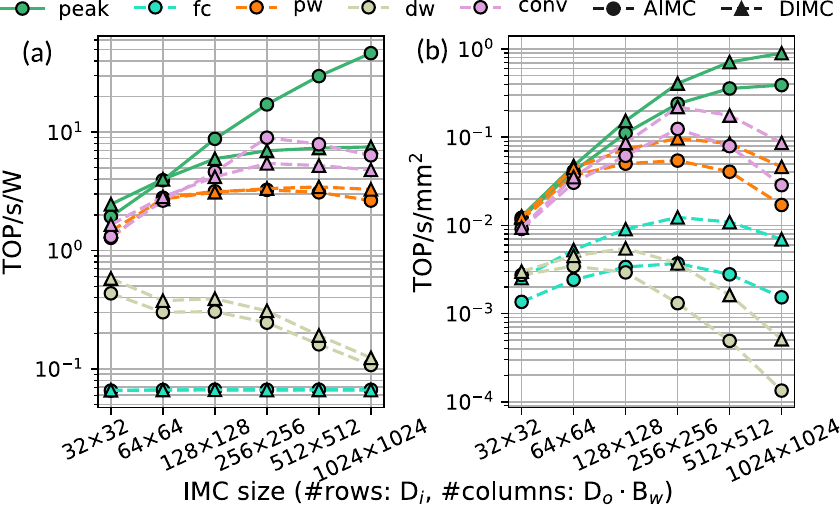}
    \caption{Single layer system performance with scaling IMC sizes}
    \label{fig:layer_topsw}
\end{figure}
%%%%%%%%%%% figure %%%%%%%%%%%%%%

%%%%%%%%%%% table %%%%%%%%%%%%%%

\input{table_layers}
%%%%%%%%%%% table %%%%%%%%%%%%%%

%%%%%%%%%%% figure %%%%%%%%%%%%%%
\begin{figure}[tbp]
    \centering
    \includegraphics[width=\linewidth]{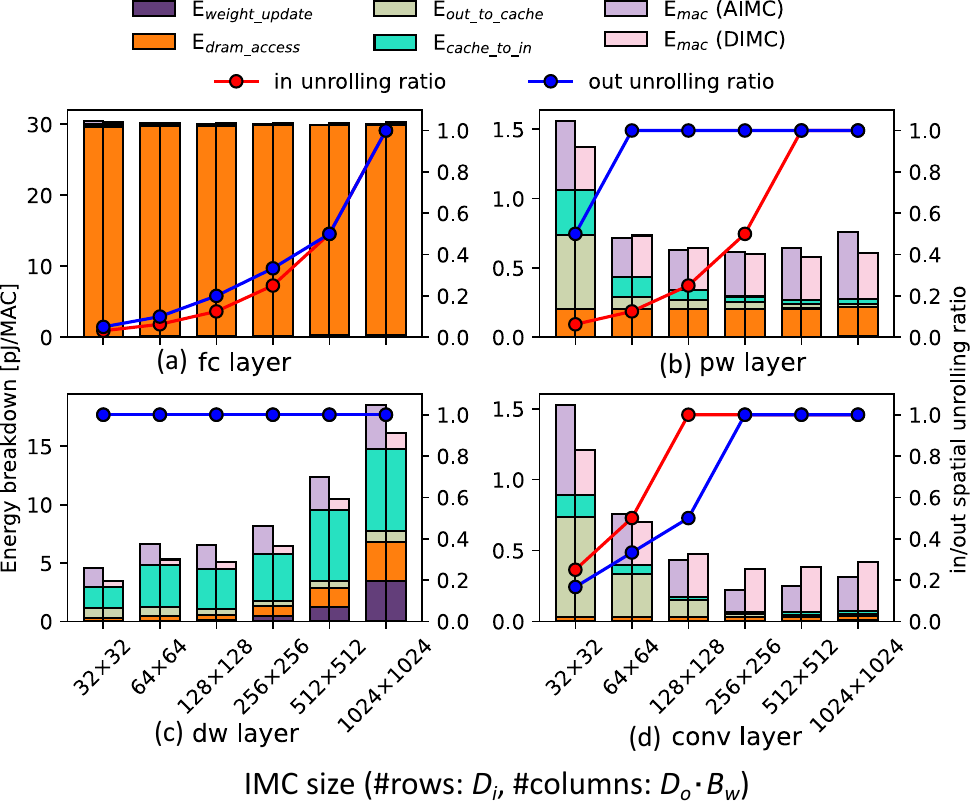}
    \caption{Single layer energy breakdown with scaling IMC sizes}
    \label{fig:layer_energy_breakdown}
\end{figure}
%%%%%%%%%%% figure %%%%%%%%%%%%%%

While for peak performance the main concern is to amortize the overhead of the peripherals, here the efficiency will strongly depend on 1) the effective workload utilization of the IMC array; 2) the amount of spatial reuse (multicasting) of the activations; and 3) the cost of loading the weights into the weight stationary array (from DRAM @3.7 pJ/bit \cite{dram_energy}), which is no longer neglected in this workload study.
All these factors depend on the applied spatial and temporal mapping. The ZigZag mapper \cite{zigzag} is utilized to optimize the deployed dataflow for each assessed architecture and workload combinations.

To dissect the actual workload efficiencies, the assessment starts by evaluating how individual layer types impact hardware efficiency.  Fig. \ref{fig:layer_topsw} compares the energy and throughput efficiencies of various layer types, specified in Table~\ref{tab:table_layer}, against peak system efficiency. 
Energy efficiency is higher for those layers that exhibit layer shapes that can best be spatially unrolled along the array dimensions and do not require frequent weight updates. This is the case for pointwise (pw) and convolutional layers (conv). Depthwise (dw) and fully connected (fc) layers, in contrast, are less well suited for IMC operation, as visualized in the energy breakdowns of Fig.~\ref{fig:layer_energy_breakdown}:  fc layers lack weight reuse, requiring a costly weight update after each MVM operation. dw layers operate with a single kernel on a single input channel, resulting in limited unrolling possibilities, even when assuming OX unrolling along the D$_o$ dimension \cite{diana}. To identify the most energy efficient array size on a layer-basis, we can look at the \textit{in/out unrolling ratio}, which indicates the fraction of the spatial reuse of the in/outputs of a layer, compared to the maximum reuse possible for that layer.
This is visualized in Fig.~\ref{fig:layer_energy_breakdown}, clearly illustrating the most efficient configuration as the first point where the maximum ratio is achieved. Larger array sizes exceed the requirement of the workload, increasing the fJ/MAC, T$_{clk}$ and area without increasing effective throughput.

%%%%%%%%%%% figure %%%%%%%%%%%%%%
\begin{figure}[tbp]
    \centering
    \includegraphics[width=\linewidth]{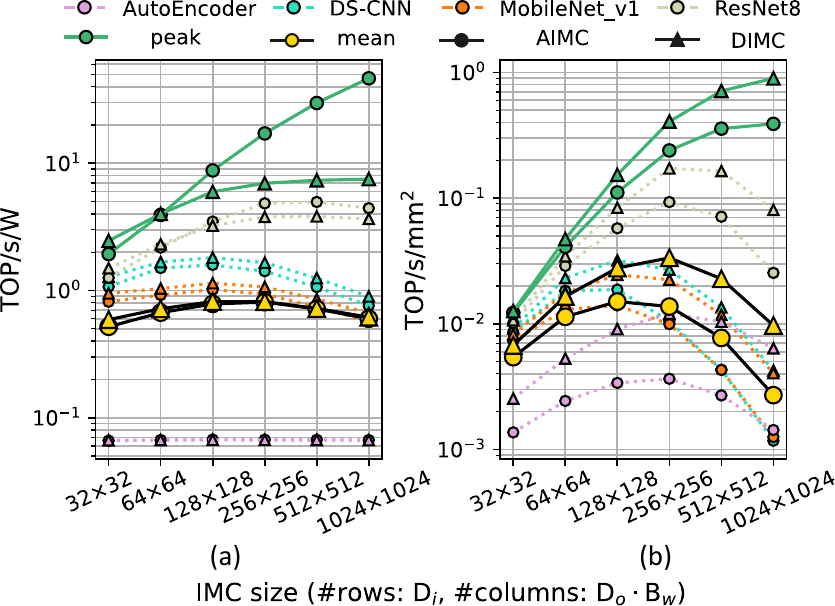}
    \caption{Workload system performance with scaling IMC sizes}
    \label{fig:workload_topsw}
\end{figure}
%%%%%%%%%%% figure %%%%%%%%%%%%%%

Finally, the evaluation is broadened to cover entire networks from the MLPerf Tiny suite, as shown in Fig. \ref{fig:workload_topsw}.
The results reflect the previous conclusions: those models that majorly consist of conv and pw layers perform better than those dominated by dw and fc layers. A distribution of the workload types for each model can be found in Fig.~\ref{fig:mlperf_workload}.
Resnet8, entirely consisting of convolutional layers, has the highest weight reuse opportunity and therefore achieves a TOP/s/W closest to the peak TOP/s/W. The fc-heavy Autoencoder suffers from frequent weight reloading, while MobileNet and DS-CNN suffer from severe macro under-utilization for big macro sizes.
Ultimately, a geometric mean of the performance across all 4 MLPerf Tiny models is provided. 
Compared to AIMC, DIMC has a higher workload TOP/s/mm$^2$ on average, while its energy efficiency is similar at the system level.

%% file: table_layers.tex
\begin{table}[tbp]
\centering

\caption{Layer parameters}
\begin{tabular}{|c|c|c|c|c|c|c|c|}
\hline

% row #1 (column title)
\multicolumn{1}{|l|}{\textbf{Type (source)}} &
  \multicolumn{1}{c|}{\textbf{$G$}} &
  \multicolumn{1}{c|}{\textbf{$OX$}} &
  \multicolumn{1}{c|}{\textbf{$OY$}} &
  \multicolumn{1}{c|}{\textbf{$K$}} &
  \multicolumn{1}{c|}{\textbf{$C$}} &
  \multicolumn{1}{c|}{\textbf{$FX$}} &
  \multicolumn{1}{c|}{\textbf{$FY$}}\\   \hline

% fc layer: from ae, layer_id: 0
\multicolumn{1}{|l|}{\text{fc (AutoEncoder)}} &
  \multicolumn{1}{c|}{\text{1}} &
  \multicolumn{1}{c|}{\text{1}} &
  \multicolumn{1}{c|}{\text{1}} &
  \multicolumn{1}{c|}{\text{128}} &
  \multicolumn{1}{c|}{\text{640}} &
  \multicolumn{1}{c|}{\text{1}} &
  \multicolumn{1}{c|}{\text{1}}\\   \hline

% pw layer: from dscnn, layer_id: 2
%\multicolumn{1}{|l|}{\text{pw (pointwise)}} &
%  \multicolumn{1}{c|}{\text{1}} &
%  \multicolumn{1}{c|}{\text{25}} &
%  \multicolumn{1}{c|}{\text{5}} &
%  \multicolumn{1}{c|}{\text{64}} &
%  \multicolumn{1}{c|}{\text{64}} &
%  \multicolumn{1}{c|}{\text{1}} &
%  \multicolumn{1}{c|}{\text{1}}\\   \hline

% pw layer: from mb, layer_id: 10
\multicolumn{1}{|l|}{\text{pw (MobileNet\_v1)}} &
  \multicolumn{1}{c|}{\text{1}} &
  \multicolumn{1}{c|}{\text{12}} &
  \multicolumn{1}{c|}{\text{12}} &
  \multicolumn{1}{c|}{\text{64}} &
  \multicolumn{1}{c|}{\text{64}} &
  \multicolumn{1}{c|}{\text{1}} &
  \multicolumn{1}{c|}{\text{1}}\\   \hline

% dw layer: from ds-cnn, layer_id: 1
\multicolumn{1}{|l|}{\text{dw (DS-CNN)}} &
  \multicolumn{1}{c|}{\text{64}} &
  \multicolumn{1}{c|}{\text{25}} &
  \multicolumn{1}{c|}{\text{5}} &
  \multicolumn{1}{c|}{\text{1}} &
  \multicolumn{1}{c|}{\text{1}} &
  \multicolumn{1}{c|}{\text{3}} &
  \multicolumn{1}{c|}{\text{3}}\\   \hline

% conv layer: from resnet8, layer_id: 2
\multicolumn{1}{|l|}{\text{conv (ResNet8)}} &
  \multicolumn{1}{c|}{\text{1}} &
  \multicolumn{1}{c|}{\text{32}} &
  \multicolumn{1}{c|}{\text{32}} &
  \multicolumn{1}{c|}{\text{16}} &
  \multicolumn{1}{c|}{\text{16}} &
  \multicolumn{1}{c|}{\text{3}} &
  \multicolumn{1}{c|}{\text{3}}\\   \hline

\end{tabular}
\label{tab:table_layer}
\end{table}

%% file: 7_conclusions.tex
\section{Conclusion}
The wide diversity in AIMC and DIMC architectures and silicon technologies published in literature makes it hard to fairly assess their optimal configurations and relative strengths. To this end, this paper proposed a modeling approach for SRAM-based IMC and conducted a comprehensive performance evaluation across various array sizes. The performance was assessed at macro level and at system level, both for peak and actual workloads.
Though sparsity impact on the energy is not considered, we clearly show the importance of evaluating IMC system-level performance on real workloads, as higher peak performance does not always lead to higher effective performance. Our model allows further quantitative comparisons across a broad set of AIMC and DIMC processor system architectures, aiding designers in making more informed decisions during the early phase of IMC architecture selection for their applications.
The derived cost models and system level benchmarking framework are provided open source at https://github.com/KULeuven-MICAS/zigzag-imc.

\section{Acknowledgements}
We want to thank Boris Murmann for his suggestions on AIMC modeling. This project has been partly funded by the European Research Council (ERC) under grant agreement No. 101088865, the European Union’s Horizon 2020 programme under grant agreement No. 101070374, the Flanders AI Research Program and KU Leuven.

%% file: 0_main.bbl
% Generated by IEEEtran.bst, version: 1.14 (2015/08/26)
\begin{thebibliography}{10}
\providecommand{\url}[1]{#1}
\csname url@samestyle\endcsname
\providecommand{\newblock}{\relax}
\providecommand{\bibinfo}[2]{#2}
\providecommand{\BIBentrySTDinterwordspacing}{\spaceskip=0pt\relax}
\providecommand{\BIBentryALTinterwordstretchfactor}{4}
\providecommand{\BIBentryALTinterwordspacing}{\spaceskip=\fontdimen2\font plus
\BIBentryALTinterwordstretchfactor\fontdimen3\font minus
  \fontdimen4\font\relax}
\providecommand{\BIBforeignlanguage}[2]{{%
\expandafter\ifx\csname l@#1\endcsname\relax
\typeout{** WARNING: IEEEtran.bst: No hyphenation pattern has been}%
\typeout{** loaded for the language `#1'. Using the pattern for}%
\typeout{** the default language instead.}%
\else
\language=\csname l@#1\endcsname
\fi
#2}}
\providecommand{\BIBdecl}{\relax}
\BIBdecl

\bibitem{murmann_aimc}
B.~Murmann, ``Mixed-signal computing for deep neural network inference,''
  \emph{IEEE Transactions on Very Large Scale Integration (VLSI) Systems},
  vol.~29, no.~1, pp. 3--13, 2021.

\bibitem{aimc1}
I.~A. Papistas, S.~Cosemans, B.~Rooseleer, J.~Doevenspeck, M.-H. Na, A.~Mallik,
  P.~Debacker, and D.~Verkest, ``A 22 nm, 1540 top/s/w, 12.1 top/s/mm2
  in-memory analog matrix-vector-multiplier for dnn acceleration,'' in
  \emph{2021 IEEE Custom Integrated Circuits Conference (CICC)}, 2021, pp.
  1--2.

\bibitem{aimc2}
J.-W. Su, Y.-C. Chou, R.~Liu, T.-W. Liu, P.-J. Lu, P.-C. Wu, Y.-L. Chung, L.-Y.
  Hong, J.-S. Ren, T.~Pan, C.-J. Jhang, W.-H. Huang, C.-H. Chien, P.-I. Mei,
  S.-H. Li, S.-S. Sheu, S.-C. Chang, W.-C. Lo, C.-I. Wu, X.~Si, C.-C. Lo, R.-S.
  Liu, C.-C. Hsieh, K.-T. Tang, and M.-F. Chang, ``A 8-b-precision 6t sram
  computing-in-memory macro using segmented-bitline charge-sharing scheme for
  ai edge chips,'' \emph{IEEE Journal of Solid-State Circuits}, vol.~58, no.~3,
  pp. 877--892, 2023.

\bibitem{aimc3}
P.~Chen, M.~Wu, W.~Zhao, J.~Cui, Z.~Wang, Y.~Zhang, Q.~Wang, J.~Ru, L.~Shen,
  T.~Jia, Y.~Ma, L.~Ye, and R.~Huang, ``7.8 a 22nm delta-sigma
  computing-in-memory ($\delta\sigma$cim) sram macro with near-zero-mean
  outputs and lsb-first adcs achieving 21.38tops/w for 8b-mac edge ai
  processing,'' in \emph{2023 IEEE International Solid- State Circuits
  Conference (ISSCC)}, 2023, pp. 140--142.

\bibitem{dimc1}
F.~Tu, Y.~Wang, Z.~Wu, L.~Liang, Y.~Ding, B.~Kim, L.~Liu, S.~Wei, Y.~Xie, and
  S.~Yin, ``A 28nm 29.2tflops/w bf16 and 36.5tops/w int8 reconfigurable digital
  cim processor with unified fp/int pipeline and bitwise in-memory booth
  multiplication for cloud deep learning acceleration,'' in \emph{2022 IEEE
  International Solid- State Circuits Conference (ISSCC)}, vol.~65, 2022, pp.
  1--3.

\bibitem{dimc2}
B.~Yan, J.-L. Hsu, P.-C. Yu, C.-C. Lee, Y.~Zhang, W.~Yue, G.~Mei, Y.~Yang,
  Y.~Yang, H.~Li, Y.~Chen, and R.~Huang, ``A 1.041-mb/mm2 27.38-tops/w
  signed-int8 dynamic-logic-based adc-less sram compute-in-memory macro in 28nm
  with reconfigurable bitwise operation for ai and embedded applications,'' in
  \emph{2022 IEEE International Solid- State Circuits Conference (ISSCC)},
  vol.~65, 2022, pp. 188--190.

\bibitem{dimc3}
A.~Guo, X.~Si, X.~Chen, F.~Dong, X.~Pu, D.~Li, Y.~Zhou, L.~Ren, Y.~Xue,
  X.~Dong, H.~Gao, Y.~Zhang, J.~Zhang, Y.~Kong, T.~Xiong, B.~Wang, H.~Cai,
  W.~Shan, and J.~Yang, ``A 28nm 64-kb 31.6-tflops/w digital-domain
  floating-point-computing-unit and double-bit 6t-sram computing-in-memory
  macro for floating-point cnns,'' in \emph{2023 IEEE International Solid-
  State Circuits Conference (ISSCC)}, 2023, pp. 128--130.

\bibitem{dimc4}
J.~Yue, C.~He, Z.~Wang, Z.~Cong, Y.~He, M.~Zhou, W.~Sun, X.~Li, C.~Dou,
  F.~Zhang, H.~Yang, Y.~Liu, and M.~Liu, ``A 28nm 16.9-300tops/w
  computing-in-memory processor supporting floating-point nn inference/training
  with intensive-cim sparse-digital architecture,'' in \emph{2023 IEEE
  International Solid- State Circuits Conference (ISSCC)}, 2023, pp. 1--3.

\bibitem{zigzag}
L.~Mei, P.~Houshmand, V.~Jain, S.~Giraldo, and M.~Verhelst, ``Zigzag: Enlarging
  joint architecture-mapping design space exploration for dnn accelerators,''
  \emph{IEEE Transactions on Computers}, vol.~70, no.~8, pp. 1160--1174, Aug
  2021.

\bibitem{imc_trends_1}
N.~R. Shanbhag and S.~K. Roy, ``Comprehending in-memory computing trends via
  proper benchmarking,'' in \emph{2022 IEEE Custom Integrated Circuits
  Conference (CICC)}, 2022, pp. 01--07.

\bibitem{imc_trends_2}
R.~Sehgal and J.~P. Kulkarni, ``Trends in analog and digital intensive
  compute-in-sram designs,'' in \emph{2021 IEEE 3rd International Conference on
  Artificial Intelligence Circuits and Systems (AICAS)}, 2021, pp. 1--4.

\bibitem{burr_benchmarking}
G.~W. Burr, S.~Lim, B.~Murmann, R.~Venkatesan, and M.~Verhelst, ``Fair and
  comprehensive benchmarking of machine learning processing chips,'' \emph{IEEE
  Design \& Test}, vol.~39, no.~3, pp. 18--27, 2022.

\bibitem{imc_trends_3}
S.~Rai, M.~Liu, A.~Gebregiorgis, D.~Bhattacharjee, K.~Chakrabarty, S.~Hamdioui,
  A.~Chattopadhyay, J.~Trommer, and A.~Kumar, ``Perspectives on emerging
  computation-in-memory paradigms,'' in \emph{2021 Design, Automation \& Test
  in Europe Conference \& Exhibition (DATE)}, 2021, pp. 1925--1934.

\bibitem{imc_trends_4}
B.~Taylor, Q.~Zheng, Z.~Li, S.~Li, and Y.~Chen, ``Processing-in-memory
  technology for machine learning: From basic to asic,'' \emph{IEEE
  Transactions on Circuits and Systems II: Express Briefs}, vol.~69, no.~6, pp.
  2598--2603, 2022.

\bibitem{9218657}
H.~Kim, Y.~Kim, S.~Ryu, and J.-J. Kim, ``Algorithm/hardware co-design for
  in-memory neural network computing with minimal peripheral circuit
  overhead,'' in \emph{2020 57th ACM/IEEE Design Automation Conference (DAC)},
  2020, pp. 1--6.

\bibitem{imc_modeling_2}
\BIBentryALTinterwordspacing
B.~Patrick, R.~Guy, R.~Nir, P.~Bruno, H.~Edward, and C.~Yiran, ``Analog,
  in-memory compute architectures for artificial intelligence,'' \emph{cs},
  vol. cs/2302.06417, 2023, arXiv: 2302.06417. [Online]. Available:
  \url{https://arxiv.org/abs/2302.06417}
\BIBentrySTDinterwordspacing

\bibitem{imc_modeling_3}
G.~Fernando, B.~Ali, V.~Kanishkan, C.~Henk, and D.~Shidhartha, ``Saca:
  System-level analog cim accelerators simulation framework: Accurate
  simulation of non-ideal components,'' in \emph{2022 37th Conference on Design
  of Circuits and Integrated Circuits (DCIS)}, 2022, pp. 01--06.

\bibitem{imc_modeling_4}
D.~Bhattacharjee, N.~Laubeuf, S.~Cosemans, I.~Papistas, A.~Maliik, P.~Debacker,
  M.~H. Na, and D.~Verkest, ``Design-technology space exploration for energy
  efficient aimc-based inference acceleration,'' in \emph{2021 IEEE
  International Symposium on Circuits and Systems (ISCAS)}, 2021, pp. 1--5.

\bibitem{imc_modeling_5}
S.~Spetalnick and A.~Raychowdhury, ``A practical design-space analysis of
  compute-in-memory with sram,'' \emph{IEEE Transactions on Circuits and
  Systems I: Regular Papers}, vol.~69, no.~4, pp. 1466--1479, 2022.

\bibitem{imc_modeling_6}
C.-J. Jhang, C.-X. Xue, J.-M. Hung, F.-C. Chang, and M.-F. Chang, ``Challenges
  and trends of sram-based computing-in-memory for ai edge devices,''
  \emph{IEEE Transactions on Circuits and Systems I: Regular Papers}, vol.~68,
  no.~5, pp. 1773--1786, 2021.

\bibitem{imc_modeling_8}
U.~Saxena, I.~Chakraborty, and K.~Roy, ``Towards adc-less compute-in-memory
  accelerators for energy efficient deep learning,'' in \emph{2022 Design,
  Automation \& Test in Europe Conference \& Exhibition (DATE)}, 2022, pp.
  624--627.

\bibitem{imc_dse_1}
S.~Yang, D.~Bhattacharjee, V.~B.~Y. Kumar, S.~Chatterjee, S.~De, P.~Debacker,
  D.~Verkest, A.~Mallik, and F.~Catthoor, ``Aero: Design space exploration
  framework for resource-constrained cnn mapping on tile-based accelerators,''
  \emph{IEEE Journal on Emerging and Selected Topics in Circuits and Systems},
  vol.~12, no.~2, pp. 508--521, 2022.

\bibitem{imc_dse_2}
X.~Peng, R.~Liu, and S.~Yu, ``Optimizing weight mapping and data flow for
  convolutional neural networks on rram based processing-in-memory
  architecture,'' in \emph{2019 IEEE International Symposium on Circuits and
  Systems (ISCAS)}, 2019, pp. 1--5.

\bibitem{timeloop}
A.~Parashar, P.~Raina, Y.~S. Shao, Y.-H. Chen, V.~A. Ying, A.~Mukkara,
  R.~Venkatesan, B.~Khailany, S.~W. Keckler, and J.~Emer, ``Timeloop: A
  systematic approach to dnn accelerator evaluation,'' in \emph{2019 IEEE
  International Symposium on Performance Analysis of Systems and Software
  (ISPASS)}, 2019, pp. 304--315.

\bibitem{maestro}
H.~Kwon, P.~Chatarasi, V.~Sarkar, T.~Krishna, M.~Pellauer, and A.~Parashar,
  ``Maestro: A data-centric approach to understand reuse, performance, and
  hardware cost of dnn mappings,'' \emph{IEEE Micro}, vol.~40, no.~3, pp.
  20--29, 2020.

\bibitem{cacti}
\BIBentryALTinterwordspacing
R.~Balasubramonian, A.~B. Kahng, N.~Muralimanohar, A.~Shafiee, and V.~Srinivas,
  ``Cacti 7: New tools for interconnect exploration in innovative off-chip
  memories,'' \emph{ACM Trans. Archit. Code Optim.}, vol.~14, no.~2, jun 2017.
  [Online]. Available: \url{https://doi.org/10.1145/3085572}
\BIBentrySTDinterwordspacing

\bibitem{9731681}
P.-C. Wu, J.-W. Su, Y.-L. Chung, L.-Y. Hong, J.-S. Ren, F.-C. Chang, Y.~Wu,
  H.-Y. Chen, C.-H. Lin, H.-M. Hsiao, S.-H. Li, S.-S. Sheu, S.-C. Chang, W.-C.
  Lo, C.-C. Lo, R.-S. Liu, C.-C. Hsieh, K.-T. Tang, C.-I. Wu, and M.-F. Chang,
  ``A 28nm 1mb time-domain computing-in-memory 6t-sram macro with a 6.6ns
  latency, 1241gops and 37.01tops/w for 8b-mac operations for edge-ai
  devices,'' in \emph{2022 IEEE International Solid- State Circuits Conference
  (ISSCC)}, vol.~65, 2022, pp. 1--3.

\bibitem{10067526}
B.~Wang, C.~Xue, Z.~Feng, Z.~Zhang, H.~Liu, L.~Ren, X.~Li, A.~Yin, T.~Xiong,
  Y.~Xue, S.~He, Y.~Kong, Y.~Zhou, A.~Guo, X.~Si, and J.~Yang, ``A 28nm
  horizontal-weight-shift and vertical-feature-shift-based separate-wl 6t-sram
  computation-in-memory unit-macro for edge depthwise neural-networks,'' in
  \emph{2023 IEEE International Solid- State Circuits Conference (ISSCC)},
  2023, pp. 134--136.

\bibitem{murmann_survey}
\BIBentryALTinterwordspacing
B.~Murmann, ``Adc performance survey 1997-2022.'' [Online]. Available:
  \url{https://web.stanford.edu/~murmann/adcsurvey.html}
\BIBentrySTDinterwordspacing

\bibitem{mlperf}
C.~Banbury, V.~J. Reddi, P.~Torelli, J.~Holleman, N.~Jeffries, C.~Kiraly,
  P.~Montino, D.~Kanter, S.~Ahmed, D.~Pau \emph{et~al.}, ``Mlperf tiny
  benchmark,'' \emph{Proceedings of the Neural Information Processing Systems
  Track on Datasets and Benchmarks}, 2021.

\bibitem{TOPSW_harmful}
V.~Sze, Y.-H. Chen, T.-J. Yang, and J.~S. Emer, ``How to evaluate deep neural
  network processors: Tops/w (alone) considered harmful,'' \emph{IEEE
  Solid-State Circuits Magazine}, vol.~12, no.~3, pp. 28--41, 2020.

\bibitem{dram_energy}
``Memory energy,'' \url{https://my.eng.utah.edu/~cs7810/pres/14-7810-02.pdf}.

\bibitem{diana}
P.~Houshmand, G.~M. Sarda, V.~Jain, K.~Ueyoshi, I.~A. Papistas, M.~Shi,
  Q.~Zheng, D.~Bhattacharjee, A.~Mallik, P.~Debacker, D.~Verkest, and
  M.~Verhelst, ``Diana: An end-to-end hybrid digital and analog neural network
  soc for the edge,'' \emph{IEEE Journal of Solid-State Circuits}, vol.~58,
  no.~1, pp. 203--215, 2023.

\end{thebibliography}
